\documentclass[prd,nofootinbib]{revtex4}

\usepackage{graphicx}
\usepackage{amssymb}

\def\ba{\begin{eqnarray}}
\def\ea{\end{eqnarray}}
\def\be{\begin{equation}}
\def\ee{\end{equation}}
\def\ben{\begin{equation} \nonumber}
\def\een{\end{equation}}
\def\baray{\begin{eqnarray*}}
\def\earay{\end{eqnarray*}}

\begin{document}

\title{
    {
        \begin{small}
        \hfill SCG-2007-02 \\
        \hfill PI-COSMO-67 \\
        \end{small}
    }
    {\bf\Large
        B-modes from Cosmic Strings
    }
}\author{Levon Pogosian}
\email{levon@sfu.ca}
\affiliation{Department of Physics, Simon Fraser University, 8888 University Drive, Burnaby, BC, V5A 1S6, Canada}
\author{Mark Wyman}
\email{mwyman@perimeterinstitute.ca}
\affiliation{Perimeter Institute for Theoretical Physics, \\ 31 Caroline St. N, Waterloo, ON, L6H 2A4, Canada}
\begin{abstract}
Detecting the parity-odd, or B-mode, polarization pattern
in the cosmic microwave background radiation due to primordial gravity waves is
considered to be the final observational key to confirming the inflationary paradigm. The search for viable models of inflation from particle physics
and string theory has (re)discovered another source for B-modes: cosmic strings.
Strings naturally generate as much vector mode perturbation as 
they do scalar, producing B-mode
polarization with a spectrum distinct from that expected from inflation itself.
In a large set of models, B-modes arising from cosmic strings 
are more prominent than those expected from primordial gravity waves. 
In light of this, we study the physical underpinnings of string-sourced B-modes
and the model dependence of the amplitude
and shape of the $C_l^{BB}$ power spectrum. Observational detection of a string-sourced B-mode spectrum would be a direct probe of post-inflationary physics near the GUT scale. 
Conversely, non-detection
 would put an upper limit on a possible cosmic
string tension of $G\mu \lesssim 10^{-7}$ within the next three years.
\end{abstract}

\maketitle
\vspace*{-0.2in}

\section{Introduction}
\vspace{-0.1in}

In the past few years, observations of the cosmic microwave background (CMB) and large scale
structure data have largely confirmed the predictions of the inflationary paradigm.
Now, observational cosmologists have begun to turn their instruments towards the final unconfirmed
signature of inflation: primordial B-mode, or odd-parity, polarization
in the CMB \cite{tfcmb}. Meanwhile, string theorists have worked diligently towards realizing observationally
viable models of inflation within reasonably well-understood
string compactifications (see, e.g.,  \cite{HenryTye:2006uv, McAllister:2007bg, Burgess:2007pz} for reviews). 
In parallel, viable models of hybrid inflation continue to be explored in the context of particle physics (e.g., \cite{smooth1,cop,rc,linde}). 
Brane and hybrid inflation models often predict the production of cosmic strings \cite{HenryTye:2006uv, kofman,tkachev,jeannerot,mairi,cop,rc, Jeannerot:2006dy,jst,costring,DV03,Polchinski03,mcgill,Firouzjahi:2005dh}, linear defects arising from the breaking of U(1) symmetries towards the end
of the inflationary epoch. These relics of the inflationary period provide
a rich phenomenology, but here we focus on 
their contribution to the polarization of the CMB. 

The production of cosmic strings in such models seems at first unimportant, since
they cannot source the majority of the CMB anisotropy \cite{ABR97}. However, it now appears that
strings may be the most prominent source of observable B-mode polarization
in many of these braneworld or hybrid inflationary models \cite{Seljak:2006hi,Battye:2007si}.
Strings, as extended objects, source vector mode perturbations in the CMB as readily as they do scalar mode perturbations. This is in contrast with the inflationary picture, where
gravity waves generated by inflation source tensor mode perturbations that are generally much
smaller than the adiabatic scalar mode perturbations.
Defect-sourced B-mode polarization was first calculated in \cite{pen_pol} for global defects,
and for local strings in \cite{battye_98, Benabed:2000tr,PTWW03}. Recently, B-mode predictions from field theoretical simulations of local strings were also reported in \cite{Bevis:2007qz}. It is true that cosmic strings are 
subdominant sources of CMB perturbation, limited to producing less than about 10\% of the
primordial anisotropy \cite{contaldi,battye,bouchet,bevis,Wyman:2005tu,Seljak:2006bg,Bevis:2007gh}\footnote{It was shown in \cite{Bevis:2007gh,Battye:2007si} that CMB data can actually favor a contribution from strings if the inflationary spectrum is either exactly Harrison-Zeldovich ($n_s=1$) or, more generally, has $n_s \ge 1$. The degree of preference is somewhat dependent on the shape of the spectrum. String models that predict a pronounced peak at $\ell \sim 400$ in the string contribution to the CMB spectrum  are less likely to improve the fit to the data than those with no peak}.
But since the inflationary tensor-to-scalar ratio - the parameter that governs
the strength of inflationary gravity waves -- is expected 
also to be below 10\%, cosmic strings and primordial
gravity waves may well be observables of similar detectability. 

Models of hybrid and brane inflation \cite{Spergel:2006hy},
tend to produce a low tensor-to-scalar ratio \cite{Martineau:2007dj}. 
Hybrid inflationary models from particle physics, for instance, typically have unobservably small ($r \lesssim 10^{-4}$)
tensor to scalar ratios \cite{Linde:1993cn, Dufaux:2007pt}. Most models of inflation that have
been realized in string theory are of the hybrid type, and also have  small scalar-to-tensor
ratios, typically $r \sim 10^{-10}$ (e.g. \cite{Brax:2006yq, Avgoustidis:2006zp, Bond:2006nc}). 
On the other hand, many of these models produce cosmic strings
near the GUT scale \cite{Lin:2006xta, Clauwens:2007wc,Jeannerot:2006jj, Cui:2007js}, 
which implies a cosmic string tension ($G\mu$) close to the observational upper bound of $G\mu \lesssim 2 \times 10^{-7}$
\cite{Wyman:2005tu,Seljak:2006bg,Bevis:2007gh}. 

Among models of inflation motivated by string theory, brane inflation 
(reviewed recently in Refs. \cite{Burgess:2007pz, HenryTye:2006uv})
can create the widest range of cosmic sting tensions, 
with  $10^{-13} < G\mu < 10^{-6}$ possible. 
In certain warped throat ``slow-roll'' scenarios, for instance,
 $G\mu \sim 10^{-9}$ is expected, while in other
``fast-roll'' realizations $G\mu \gtrsim 10^{-7}$ can easily be achieved \cite{Firouzjahi:2005dh}.
In the low-multipole region ($2 < \ell < 100$), experimental sensitivity to primordial
gravity waves can be converted to sensitivity to strings via\footnote{This conversion covers only the region of $2 < \ell < 100$, and was
calculated by comparing the total BB power in strings versus the total BB power
from tensor modes in this $\ell$ range. This particular relation is dependent on the parameters of the string model, and a more complete conversion is suggested in  \S\ref{ddetection}.} 
$G\mu\leftrightarrow 1.4 \times 10^{-6}\sqrt{r}$; hence,
the promise that near-term experiments can observe $r\sim10^{-2}$ 
in this multipole range implies sensitivity
to strings with tensions as low as $G\mu \sim 10^{-7}$. 
Brane inflation, however, may not be able \footnote{More complicated set-ups involving ``wrapped'' branes might loosen such bounds \cite{Krause:2007jr, Becker:2007ui}.} to create a tensor-to-scalar ratio above $r \sim \mbox{(few)} \times 10^{-3}$ \cite{Baumann:2006cd}. For high $\ell > 100$, the 
situation is different: the dominant B-modes are expected to come from the lensing of
primordial-E modes into B-modes by large scale structure. However, the string sourced
B-mode spectrum peaks in this high $\ell$ range, so if strings are present they will
show up as an excess over the expected signal. An apparent excess of 10\%, for instance,
would be caused by strings with a tension of $G\mu \simeq 6 \times 10^{-8}$ (assuming
standard string network parameters).

To summarize: in models of inflation that are motivated either by string theory or particle physics, the more promising avenue for creation of B-mode polarization is via cosmic strings, not tensor modes.
Hence, we might hope that if B-modes from inflationary tensor modes
are not observed, B-modes from cosmic strings produced after inflation might be. 
Inflationary models with large tensor modes
(like chaotic inflation) tend not to create defects. Distinguishing
the source of the polarization will be a challenge. To distinguish
string B-modes from primordial gravity waves requires fine
scale resolution in the range $\ell < 100$. Distinguishing string sourced
polarization at $\ell > 100$ requires an accurate calculation of the amplitude of 
the B-mode polarization that comes
from the conversion of E-mode polarization to B-mode polarization by gravitational lensing.
This is because the 
peak BB contribution from strings falls at $\ell \sim 1000$, overlapping heavily with
the spectrum expected from lensing of E. Hence, the presence of a string-induced B-mode 
would induce an excess in power throughout this region of $\ell$ values
above what is expected from lensing. Accurately characterizing this excess will
require very good predictions of the expected lensing signal. 
If all goes well, polarization experiments could eventually see strings with
 tensions as low as $G\mu \sim 10^{-9}$ \cite{Seljak:2006hi}.
Because strings can source B-modes in the aftermath of a hybrid inflationary
epoch that cannot directly source B-modes, the class of inflationary models that 
B-mode experiments can test is broadened. Detecting neither string or primordial 
gravity wave sourced B-modes would strongly constrain many 
inflationary models. These constraints would be different from, and perhaps
stronger than, the constraints derived from measurement of other cosmological
parameters (e.g., the spectral index $n_s$).

In this paper, we outline the physics behind cosmic strings generation of 
B-mode polarization, placing a heavy emphasis on how much variability might be expected
from various models of string networks. This gives us an idea of how much a successful detection
of cosmic string B-modes might teach us about these networks.
 We also review some of the currently operating, soon to be
launched, and future planned experiments to measure the CMB's B-mode polarization to 
give estimates for what range of string tensions they will be able to detect.

\section{How Strings Polarize the CMB}\label{polarization}
\vspace{-0.1in}

The string-sourced perturbations are fundamentally different from those produced by inflation. Inflation sets the initial conditions for the perturbations (i.e. their initial spectrum) which then evolve in time without additional disturbances being produced. In this sense, inflationary perturbations can be classified as {\it passive}. Inhomogeneities produced by cosmic strings are {\it active} -- string networks are thought to persist throughout the history of the universe and actively source metric perturbations at all times. It is well known \cite{svt} that, in the absence of a source term, vector modes decay quickly, scalar modes have a growing and a decaying solution, and tensor modes are sustained on superhorizon scales and decay after entering the horizon. For this reason, vector modes are not relevant in passive scenarios, and are rarely considered in the literature. Cosmic strings, on the other hand, actively source scalar, vector, and tensor perturbations. For them, scalar and vector mode perturbations are typically of similar magnitude. The string generated tensor modes are also at a comparable level, but their observational impact is generally lower because of the oscillatory nature of gravity waves \cite{Turok:1997gj}.

The non-zero CMB temperature quadrupole at the time of recombination leads to linear polarization of CMB through Compton scattering of photons on baryons. Before recombination, the photons and baryons form a tightly coupled fluid with the anisotropy characterized only by a monopole (density contrast) and a dipole (mass flow) components. After
last scattering, photons propagate freely. Hence polarization is produced during a narrow time window in the last scattering epoch, when scattering become sufficiently rare for a temperature quadrupole to develop. Essentially, this window corresponds to the time between the last scattering and the next-to-last scattering. Polarization can also be produced at more recent epochs during reionization.

At any point on the sky, CMB polarization can be represented by a headless vector with a magnitude $|P|\equiv \sqrt{Q^2+U^2}$ and an orientation angle $\nu \equiv (1/2) \arctan(U/Q)$, 
where $Q$ and $U$ are two of the four Stokes' parameters \cite{stokes}\footnote{The other two parameters are the intensity $I$, and the one quantifying the circular polarization, $V$. Circular polarization is not produced at last scattering and is expected to be zero on cosmological scales in the absence of large scale magnetic fields or other exotic physics.}. The pattern of the $P$-vectors on the sky can be separated into parity-even and parity-odd contributions, or the so-called $E$ and $B$ modes. While intensity gradients automatically generate $E$-mode patterns, the $B$-mode is not produced unless there are metric perturbations that can locally have a non-vanishing handedness, i.e. have a parity-odd component. Local departures from zero handedness can be due to tensor modes, i.~e. gravity waves, which can be represented as linear combinations of left- and right-handed polarizations \cite{carroll_book}. The B-mode can arise from non-zero vector modes, or vorticity.

Cosmic strings induce both vector and tensor components in the metric by actively generating
anisotropic stress, with the vector contribution being the most important.  This implies a non-vanishing vorticity in the photon-baryon plasma at the surface of last scattering, which generates B-mode polarization in the CMB.  The vector metric potential $V$ in Fourier space, in the generalized Newtonian gauge, can be defined as \cite{Hu:1997hp}
\be
h_{0j}({\mathbf k},\eta)e^{i{\mathbf k}{\mathbf x}}=-V Q^{(1)} \ ,
\ee
where $h_{\mu\nu}$ is the linear perturbation to the metric and $Q^{(1)}$ is a divergenceless harmonic eigenmode
of the Laplacian. The reader is referred to \cite{Hu:1997hp} and \cite{our_erratum} for more detailed description of the formalism. At last scattering, $V$ is given by \cite{Hu:1997hp},
\be
V(\eta_*, k) = - 8 \pi G a^{-2}_* \int_0^{\eta_*} d \eta \, a^4 \, k^{-1} \, (p_f \pi_f^{(1)} + \pi_{s}^{(1)} )\ ,
\ee
where $f$ represents the fluid and $s$ the seed contributions to the anisotropic stress, a subscript $*$ indicates the values of the conformal time $\eta$ and scale factor $a$ at last scattering.
The vector anisotropic stress of the fluid, included here, is indirectly sourced by the strings through the metric. 
However, the direct coupling of the string anisotropic stress to the vector metric potential dominates over the (indirectly sourced) fluid contribution. Finally, one can see from this expression that the vector potential will decay as $a^{-2}$ if it is not actively seeded.

It is instructive to consider the vector mode contribution from one straight cosmic string segment.
Without loss of generality, one can choose ${\mathbf k}=\hat{\mathbf z}k$ \cite{ABR99}, in which case the $T_{13}$ and $T_{23}$ components of the Fourier transform of the energy-momentum tensor of the string network become pure vector modes, i.e. they do not contribute to the scalar or tensor modes. The two components are statistically equivalent, and it is sufficient to work with just one of them. e.g. $T_{13}$. The Fourier components of the energy-momentum tensor of the string are given by \cite{ABR97,ABR99, Pogosian:1999np} 
\ba
\nonumber
T_{00} & = & \frac{\mu \alpha}{\sqrt{1 - v^2}} 
\frac{\sin(k \hat{X}_3' l / 2 ) }{k \hat{X}_3' /2} \cos(\vec{k}\cdot\vec{x}_0 + k \hat{\dot X}_3 v \eta ), \\ 
T_{ij} & = &  \left [ v^2 \hat{\dot X}_i \hat{\dot X}_j - \frac{(1-v^2)}{\alpha^2} \hat{X}_i' \hat{X}_j'  \right ]   T_{00} \ ,
\label{stringstress}
\ea
where $\hat{X}_i'$ is the component of the string's orientation in the direction $i$ and $\hat{\dot X}_i$ is the string's velocity in the direction $i$, $v$ is the velocity, $\alpha$ is the wiggliness parameter that renormalizes the string's tension based on its small scale structure, 
$\mu$ is the tension, and $\vec{x}_o$ is the string's location in space. We chose ${\mathbf k}=\hat{\mathbf z}k$.  In our notation, the anisotropic stress is related to $T_{13}$ via \cite{our_erratum}
\be
\pi^{(1)}_s({\mathbf k},\eta) = -2\sqrt{2} T_{13}({\mathbf k},\eta) \ .
\label{define_emtv}
\ee
The key observation one can make from these equations is that the string contribution to anisotropic stress is always present, even for a stationary string. The string's orientation inherently violates isotropy. 

For a single string, the strength of the scalar, vector and tensor contributions depend on the orientation of the string with respect to ${\mathbf k}$. A network of strings can be thought of as a collection of many string segments at random positions, each with a random orientation and velocity. The B-mode power spectrum is determined by the two-point correlation $\langle \sum_m \pi^{(1)}_{s(m)} \sum_n \pi^{(1)*}_{s(n)} \rangle$, where the sums are taken over all segments present at last scattering. If the segments are statistically independent -- which is approximately true since they are taken to be comparable to the horizon size -- then the ensemble average reduces to a single sum over contributions of separate strings, i.e. $\langle \sum_n |\pi^{(1)}_{s(n)}|^2 \rangle$. In fact, if all strings are statistically equivalent, this reduces to $N \langle |\pi^{(1)}_{s(1)}|^2 \rangle$, where N is the number of segments and $\pi^{(1)}_s(1)$ is the contribution from one string. One can analytically evaluate the average $\langle |\pi^{(1)}_{s(1)}|^2 \rangle$ for a straight string using eqs.~(\ref{stringstress}) and (\ref{define_emtv}) by averaging over uniformly distributed velocity directions, string positions and orientations. The dependence of the B-mode spectrum on the network parameters can be inferred from how $\langle |\pi^{(1)}_{s(1)}|^2 \rangle$  varies with $v$, $l$ and $\alpha$. We will return to this when we interpret our numerical results, but we sketch the main dependencies now.

\begin{figure}[htbp] 
   \centering
   \includegraphics[width=4in]{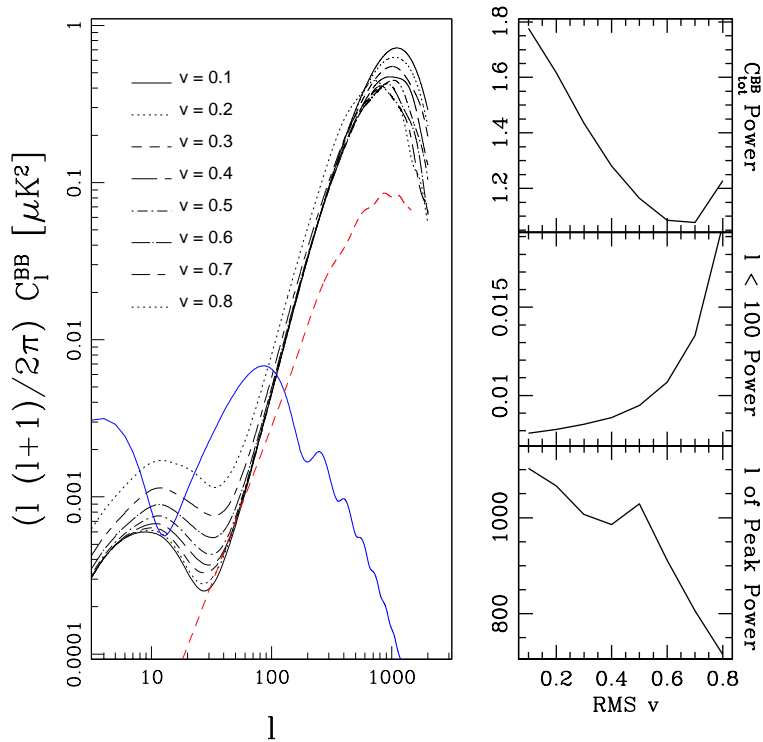} 
   \caption{Here we focus on how detectability varies with the string segments' rms velocity.
   We do not
   enforce that the TT power sourced by strings is 10\% of the total CMB TT power. Instead,
   we fix $G\mu (0.3/\xi) = 3 \times 10^{-7}$. On the left,
   we plot the spectrum, with the overall amplitude normalization factored out (see label on y-axis).
   On the right, we plot (Top) total BB-power (Middle) BB-power summed over $\ell<100$ and 
   (Bottom) the location of the peak of the spectrum. Higher low-$\ell$ power means a stronger
   signal in the gravity wave region. A lower peak $\ell$ makes it easier to distinguish string-sourced
   B-modes from the E to B lensing signal. Of note is the quadratic dependence of low-$\ell (<100)$
   power on the strings' rms velocity.}
   \label{veldouble}
\end{figure}

Two important network parameters -- the correlation length, $l$, and the strings' root-mean square (rms) velocity, $v$ -- enter $T_{00}$ (see  eq.~(\ref{stringstress})) in a sine and cosine function, respectively. Scaling $l$ and $v$ up and down, then, amounts
to changing the Fourier space dependence of the string contribution. The dominant Fourier modes will determine the dominant angular modes of the B-mode spectrum and set the location of the angular peaks. The sine function that depends on $l$ sets the location of the peak in string-sourced power. The cosine function that depends on $v$ also takes a contribution from a random phase. Varying $v$ changes the number of cosine-peaks that contribute to the shape of the overall peak. Higher velocity strings will contribute more cosine peaks and lead to broader B-mode spectra. Changing the rms
velocity also changes the ratio $\langle |T_{ij}|^2 \rangle / \langle |T_{00}|^2 \rangle$, i.~e. the relative contribution of the anisotropic stress versus the isotropic
stress. Namely, increasing $v$ {\it reduces} this ratio. This can be seen by computing this ratio from eq.~(\ref{stringstress}), 
$$
\frac{\langle |T_{ij}|^2 \rangle}{\langle |T_{00}|^2 \rangle}  = \langle |\hat{X}_i' \hat{X}_j'|^2 \rangle - 2 v^2 [ \langle |\hat{X}_i' \hat{X}_j'|^2 \rangle + 
 \langle (\hat{X}_i' \hat{X}_j' ) ( \hat{\dot X}_l \hat{\dot X}_k ) \rangle ] + \mathcal{O}(v^4),
$$
so that any nonzero velocity always decreases the anisotropic contribution.
String wiggliness similarly suppresses the anisotropic stress relative to the isotropic stress.  All of these effects can be seen in \S \ref{numerical} and Figs. \ref{sixspectra}, \ref{xidouble}, and \ref{veldouble}, which present
the results from our numerical experiments. 

\section{Dependence of the B-mode spectrum on the string network properties}

\subsection{The string network model}
\vspace{-0.1in}


\begin{figure*}[*t] 
   \centering
   \includegraphics[width=4in]{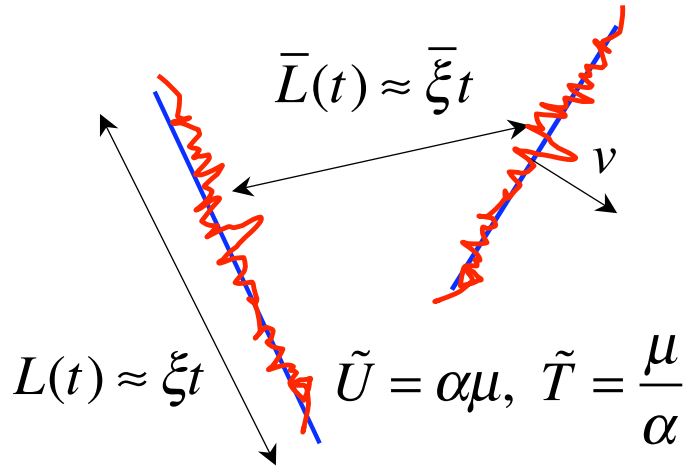} 
   \caption{Illustration of model length scale parameters.}
   \label{lengthscales}
\end{figure*}

The cosmic string model we use was introduced in \cite{ABR97,ABR99}, based on the approach suggested in \cite{hindmarsh97}, and developed into its present form in \cite{Pogosian:1999np,gangui}. The code
that evaluates the CMB temperature and polarization spectra for the model is 
publicly available as CMBACT \cite{cmbact}. In this
model, the string network is represented by a collection of
uncorrelated straight string segments
moving with uncorrelated, random velocities. There are two
fundamental length scales in such a model, as illustrated
in Fig. \ref{lengthscales}: $\xi$, the length 
of a string segment, which represents the typical length
of roughly straight segments in a full network; and
$\bar{\xi}$, the typical length between two string segments,
which sets the number density of strings in a given volume ($N_s\propto1/\bar{\xi}^2$).
The model assumes network scaling, which is 
expected from numerical simulations. In the simplest scaling, the length of each
segment at any time ($\xi t$) is taken to be equal to the correlation length
of the network ($\bar{\xi} t$). This length and the rms velocity of segments
are computed from the velocity-dependent one-scale model of
Martins and Shellard \cite{MS96}. The positions and orientations of the segments are
drawn from uniform distributions in space and on a two-sphere, respectively.
The model's parameters have been calibrated to produce source correlation functions
in agreement with those in ref. \cite{vincent}. The {\it shapes} of the spectra obtained using the model are in good agreement with results of other groups
\cite{Bevis:2007qz,magetal96,carlo99,dani}, who used different methods.

On the cosmological scales probed by the CMB measurements, the
fine details of the string evolution do not play a major role. It is
the large-scale properties -- such as the scaling distance, the
equation of state (wiggliness), and the root-mean-square (rms) velocity  -- that 
determine the shape of the string-induced spectra.
All of these effects are accounted for by, and are adjustable in, our model.
Furthermore, the overall normalization of the spectrum has a simple dependence
on the string tension and number density:
\be
C_\ell^{\rm strings} \propto N_s (G\mu)^2 \propto \left ( \frac{G\mu}{\bar{\xi}} \right)^2.
\ee
Since we can understand the overall amplitude of the spectrum
in this simple way, we will focus on how the shape and distribution
of B-mode power spectra are altered by changing various network parameters.

Finally, the wiggly nature of strings is accounted for by modifying the string
energy-momentum tensor using the wiggly string
equation of state:
\be 
\label{stringeos}
\tilde U = \alpha \mu \ , \
\tilde T = \alpha^{-1} \mu  \ ,\ee
where $\alpha$ is a parameter describing the wiggliness,
$\tilde U$ and $\tilde T$
are the mass per unit length and the string tension of the wiggly
string, and $\mu$ is the tension (or, equivalently, the mass per unit length) of 
the smooth string. In
addition to modifying the equation of state, the presence of small-scale 
structure slows strings down on large scales. We account for this by dividing 
the rms string velocity by the parameter $\alpha$.
The wiggliness of the strings remains approximately constant during the
radiation and matter eras, but changes its value during the transition between the two.
We take the radiation era value, $\alpha_r$, to be a free parameter that we vary,
and set the matter era value to be $\alpha_m=(1+\alpha_r)/2$, with a smooth
interpolation between the two values (as prescribed in \cite{Pogosian:1999np}).
For Nambu-Goto strings, this roughly agrees
with results of numerical simulations \cite{BenBouch,AllenShellard90} which show a decrease
from $\alpha_r \sim 1.8-1.9$ in the radiation era to $\alpha_m \sim 1.4-1.6$ in the matter era.

\subsection{Spectrum Dependence on Network Parameters}\label{numerical}
\vspace{-0.1in}

\begin{figure}[t!] 
   \centering
   \includegraphics[width=4.5in]{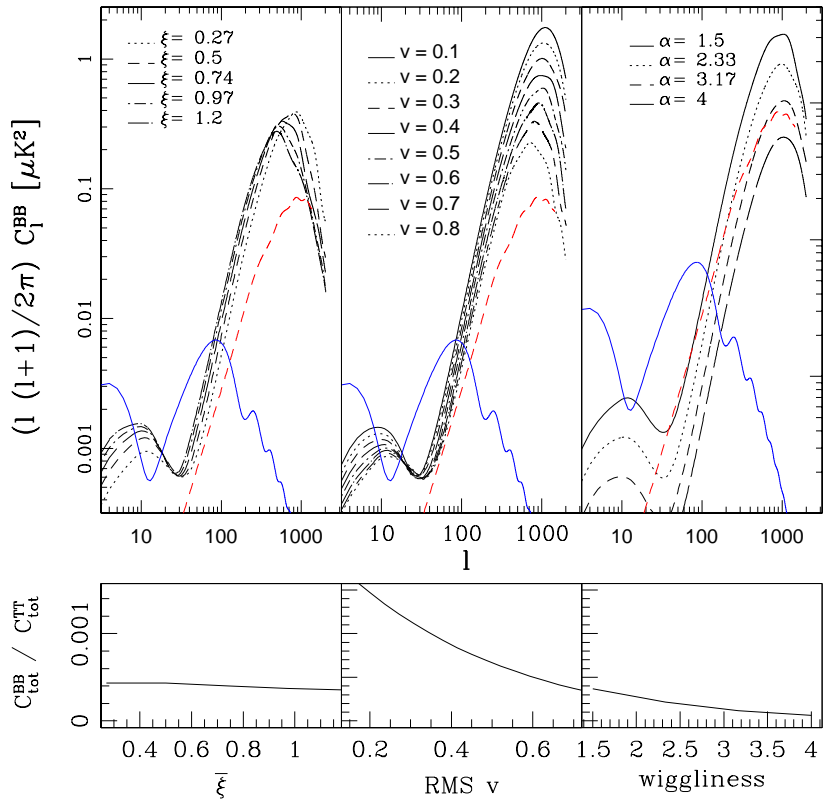} 
   \caption{In these plots, the cosmic string BB power spectrum has been normalized such that the 
   cosmic string contribution to the total CMB  TT power is 10\%. 
   (Top left) Variation of the segment correlation length (Top center) Variation of
   the string segments' rms velocity (Top right) Variation of the string wiggliness, $\alpha$.
 Note also that we have plotted  the B-mode spectra from primordial gravity waves for a tensor-to-scalar ratio $r = 0.1$ (blue, solid) and from E to B lensing (red, dashed)
   in each panel for comparison purposes.
      The lower three panels show the ratio of the BB power to TT power over the range
   of parameters considered.}
   \label{sixspectra}
\end{figure}

Before proceeding to discuss specific dependencies, let us describe the main features in the shape of the string induced B-mode spectrum. All spectra in Fig.~\ref{sixspectra} have two peaks. 
The less prominent peak at lower $\ell$ is due to rescattering of photons during reionization, which corresponds to an optical depth of $0.09$ \cite{wmap3}, or an approximate redshift range of $7 < z  < 12$. The main peak, at higher $\ell$, is the contribution from last scattering. Both peaks are quite broad.
This is because a string network seeds fluctuations over a wide range of scales  at any given time. The position of the main peak is determined by the most dominant Fourier mode stimulated at last scattering. One can estimate this dominant scale using simple analytical considerations based on the uncorrelated segment picture presented in the previous section. It primarily depends on the string correlation length at last scattering, but also is affected by the rms string velocity 
and the wiggliness. Hence, finding the location of the main string peak, which falls in the range of many B-mode experiments (see Sec. \ref{detection}), would give us a direct measure of the physics of a cosmic string network. 

We discuss the key dependencies in a sequence of subsections below. 
For fixing the normalization, there are two possible approaches. In the first,
we make use of the observational upper bound that at most $\sim10\%$
of the overall TT power is sourced by strings.
We define the total power as
\be
C^X \equiv \sum_{\ell=2}^{2000} (2\ell+1) C_\ell^X \ ,
\ee
where the superscript $X$ labels a particular type of the spectrum. The string tension, $G\mu$, controls the overall amplitude of the power spectrum, but does not affect its shape or the ratio of the BB to the TT spectrum. So we can adjust the string spectrum normalization (i.~e. $(G\mu)^2$) so that 
\be
C^{TT}_{strings}/C^{TT}_{total}=0.1 \ .
\ee 
We can then use the adjusted value of $G\mu$ to evaluate the B-mode spectrum. 
The results from this approach are shown in Fig.~\ref{sixspectra}.
Following from this approach, when we speak of the ``amplitude'' of BB power, this amplitude should be interpreted as amplitude of BB relative to TT power. In particular, in the lower half of Fig.~\ref{sixspectra} 
we plot $C^{BB}/C^{TT}$. Increasing the wiggliness of the strings, for example,
increases the TT power but does not affect the anisotropic stress. Hence, after we reduce $G\mu$ to keep the string induced CMB anisotropy at a 10\% level, the amplitude of the predicted B-mode signal will be lower for larger 
wiggliness. 

The other approach we employ is to ignore constraints on the TT spectrum and simply ask
what $G\mu$ will be detectable, given a certain experimental sensitivity to B-mode power.
 This way of looking 
at the problem allows us to neglect questions about what the exact upper limit of the
string contribution to the TT power is and gives a more direct measure of how varying
network parameters can make string-sourced BB power more or less detectable.

\subsubsection{Correlation length}
\vspace{-0.1in}

\begin{figure}[htbp] 
   \centering
   \includegraphics[width=4in]{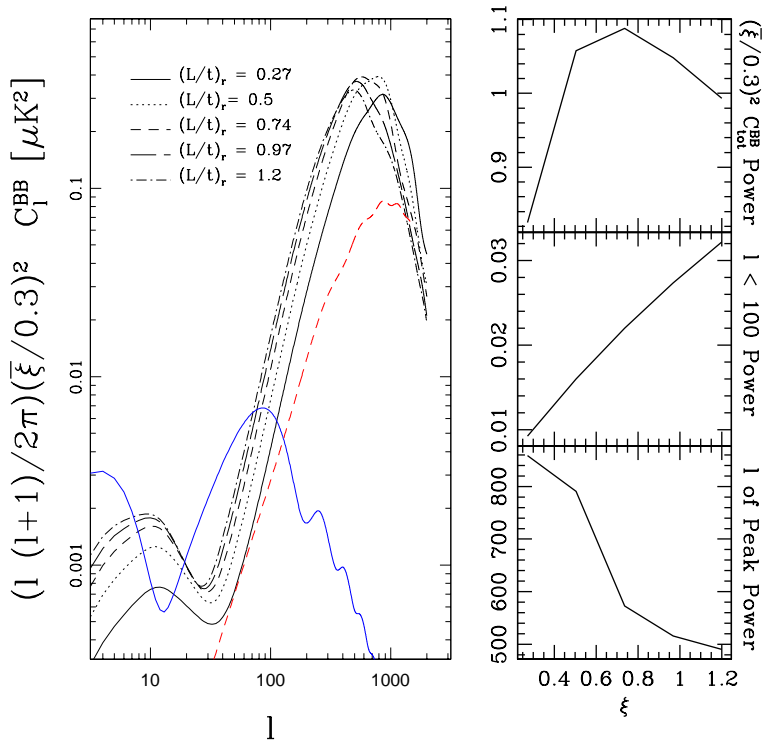} 
   \caption{Here we focus on how detectability varies with the correlation length. We do not
   enforce that the TT power sourced by strings is 10\% of the total CMB TT power. Instead
   we fix $G\mu = 3 \times 10^{-7}$. On the left,
   we plot the spectrum, with the overall amplitude normalization factored out (see label on y-axis).
   On the right, we plot (Top) total BB-power (Middle) BB-power summed over $\ell<100$ and 
   (Bottom) the location of the peak of the spectrum. Higher low-$\ell$ power means a stronger
   signal in the gravity wave region. A lower peak $\ell$ makes it easier to distinguish string-sourced
   B-modes from the E to B lensing signal. Of note here is the nearly linear dependence of
   low-$\ell (< 100)$ power on correlation length.}
   \label{xidouble}
\end{figure}

Correlation length controls the dominant momentum mode contributed by strings to the anisotropic and ordinary stress. In our string network model, the comoving length $l$ is a function of time determined from the one-scale model of \cite{MS96}. Scaling implies that the physical correlation length $L(t)=a l$ remains a fixed fraction of the horizon. That is, $L=\xi_{r,m} t$, where $\xi_r$ and $\xi_m$ are constants corresponding to the radiation and matter eras respectively. Typically, $\xi_m \approx 2 \xi_r$, and the transition between the two scaling regimes carries well into the matter era, so the value of $\xi$ during the last scattering epoch is much closer to $\xi_r$ than it is to $\xi_m$. 

To study how the B-mode spectrum changes when the correlation length varies, we changed the function $l$ by constant factors. In other words, in each case we did not alter the parameters of the one-scale model --
which encodes a relationship between $v$ and $l$ --  but simply multiplied the output length scale by a constant
before feeding it into the CMB calculations. In Figs.~\ref{sixspectra} and \ref{xidouble} we show results for six such trials, corresponding to $\xi_r=$\{0.04, 0.27, 0.5, 0.74, 0.97 and 1.2\}.

The string length enters via $\sin(Akl)$ in eq.~(\ref{stringstress}), where $A< 0.5$ is a randomized constant. Hence we would expect that the dominant string contribution -- the location of the main string power spectrum peak -- scales monotonically with $l$. A larger $l$ causes a ``longer'' mode to be dominant with the peak at a smaller $\ell$ multipole. Indeed, this is what we observe numerically, as shown in Figs.~\ref{sixspectra} and \ref{xidouble}.

The amplitude of the BB contribution relative to the TT contribution is not strongly dependent on $l$, as evident from the bottom panel in Fig.~\ref{sixspectra}. The overall amplitude of the BB contribution
is accurately captured by the $\xi^{-2}$ scaling of the string number density. Aside from this
overall rescaling, the main effect of larger string correlation length is to move the
location of the peak of the spectrum to lower $\ell$, enhancing the B-mode power
at low-$\ell < 100$, the region of $ell$ space in which strings and primordial tensor modes produce comparable power.


\subsubsection{String velocity} 
\vspace{-0.1in}

The rms string velocity remains approximately constant through the radiation era and the early matter era, with a typical value of $v_r \approx 0.65$ for the default parameters used in \cite{MS96}. To study the effect of different $v$'s on the B-mode spectrum, we employed a method similar to the one used in the previous subsection. Namely,  we did not alter the parameters of the one-scale model, but instead multiplied the output for the rms string velocity by a constant factor to obtain values of $v$ during the radiation epoch in the range $0.1 < v_r < 0.8$. 
 
As anticipated from analytical considerations in \S \ref{polarization}, larger values of the rms velocity decrease the amount of B-mode power relative to TT power. However, the dependence of the shape of the spectrum on velocity is non-trivial. The peak moves to higher multipoles 
(smaller scales) for low and moderate velocities ($v \sim 0.5$ and $v \rightarrow 0$), but moves to larger scales for higher
velocities and between the low and moderate velocities. These changes reflect the nonlinear dependence
on velocity encoded in Eq. \ref{stringstress}, averaged over an entire network. The 
chief cause of this effect seems to come from the cosine dependence of the anisotropic 
stress on velocity. As velocity grows larger, more cosine maxima contribute to the power. 
Hence the peak is not as well defined by the overall $\ell$-dependence, and becomes a 
somewhat broader peak, with many velocity-driven momentum modes contributing.

For detectability, the chief result of changing the string segments' rms velocity is that
higher velocities generally move the peak of the spectrum to lower $\ell$ and, consequently, 
force proportionally more of the B-mode power into the low-$\ell (<100)$ region. 
We should not be surprised that the effect of higher velocity is qualitatively
similar to that of larger $\xi$. The worldsheet area swept out be a string segment
is proportional to $v_{rms} \xi t$, and the effect strings have on their surroundings is related
to this area. Hence, higher velocities have a similar effect to longer correlation lengths.



\subsubsection{Wiggliness}
\vspace{-0.1in}

The amount of small-scale structure on strings, or their wiggliness, affects the B-mode spectra in two ways. One effect is that wigglier strings are slower on horizon scales, hence the trends associated with varying the rms velocity would again be relevant. The other effect is the suppression of the stress-components of the energy-momentum tensor with respect to the energy density. This follows immediately from the modified string equation of state (\ref{stringeos}) and, equivalently, from the $\alpha^{-2}$ factor in (\ref{stringstress}). Generally, wigglier strings generate less vector-mode power relative to the total power, because of the overall suppression of the anisotropic stress. The dependence of the peak location on the wiggliness comes through the effect on the rms velocity of the network. For the range of $\alpha_r$ we have considered in this work ($1<\alpha_r<4$), the effect on the location of the peak is relatively minor.

\section{Detection thresholds} \label{ddetection}
\vspace{-0.1in}

The standard inflationary paradigm predicts two sources for B-mode polarization: primordial
gravity waves and gravitational lensing of E-mode polarization into B-mode polarization
by intervening structure. The physics underlying the latter of these sources is well established,
and the spectrum and amplitude of this source of B-mode polarization can be predicted
using knowledge of the E-mode spectrum, which is generated directly from the well understood
adiabatic, scalar perturbations from inflation. B-modes from gravity waves, however, are
quite model dependent in their amplitude. Gravity waves are tensor mode sources of perturbation
that contribute to the temperature anisotropies primarily at low multipoles. Because of this,
it has become standard practice in the literature to characterize the size of the gravity wave
contribution to the primordial anisotropies through a simple model-independent parameter,
$$
r = \frac{C_{\ell = 2}^{TT} (\mbox{tensor})}{C_{\ell =2}^{TT} (\mbox{scalar})}.
$$
As mentioned in the Introduction, this ratio can reach values as high as $r\simeq 0.5$ in some large-field
models of inflation, but is typically low in hybrid and brane inflation models. The current limits 
from {\sf WMAP3} are $r < 0.33$ at 95\%c.l. \cite{wmap3}

With a large number of experiments hoping to measure CMB polarization 
either underway or in planning \cite{list}, we cannot easily 
give an exhaustive account of the ability of each to detect cosmic string sourced B-modes.
 Nonetheless, it is instructive to compare a few 
relatively near-term experiments, which give us some idea about what kind of detection thresholds will be achieved in the next couple of years. Similar reviews have been conducted in Refs. \cite{Seljak:2006hi, Bevis:2007qz}. 
Experimental targets can roughly be divided into ``high'' and ``low'' $\ell$ sensitivity ranges. The ``low''
range is roughly $\ell \leq 100$, where gravity wave-sourced B-modes would dominate if they exist.
The ``high'' range is $100 < \ell < 1500$, the range of multipoles where E to B lensing is expected
to dominate; this is also the range where the peak of string-sourced B-modes lies.

To make estimates of experimental sensitivity in the ``low'' range, we devised a simple factor for translating between gravity wave and cosmic string-sourced $C_\ell^{BB}$ power. 
Most experimental groups estimate what $r$ they will be able to reach. 
This estimate amounts to reporting their sensitivity to B-mode power in the ``low'' range.
Since strings also make B-mode power in this range, albeit with a different spectrum,
we can calculate what $G\mu$ would be necessary to produce the same B-mode
power as gravity waves for a given tensor-to-scalar ratio, $r$. 
Practically, we convert from $r$ to $G\mu$ as follows: we
generated the $C_\ell^{BB}$ spectrum using {\sf CMBFAST} for a particular $r$. We then summed the total power in the
$C_\ell^{BB}$s between $2 < \ell \leq 100$. We then used our own code to repeat the
calculation for string-sourced  $C_\ell^{BB}$s for our fiducial $G\mu$.
 Then, since $C_\ell^{BB} \propto r$ for 
tensor modes and  $C_\ell^{BB} \propto (G\mu)^2$ for cosmic strings generally,
while $C_\ell^{BB} \propto (\xi/0.3) (v_{rms}/0.65)^2$ in the low-$\ell$ region,  we were able to infer
a ``translation'' equation -- the string tension $G\mu$ necessary to create the same BB power between $2 < \ell \leq 100$ for a given scalar-to-tensor ratio, $r$, given a set of network parameters.
For our string model this is given approximately by
$$
G\mu  \leftrightarrow 1.4 \times 10^{-6} \sqrt{r} \left ( \frac{0.65}{v_{rms}}\right ) \sqrt{\frac{\xi}{0.3}} \alpha \quad \quad 2 < \ell < 100.
$$

\begin{figure}[htbp] 
   \centering
   \includegraphics[width=3in]{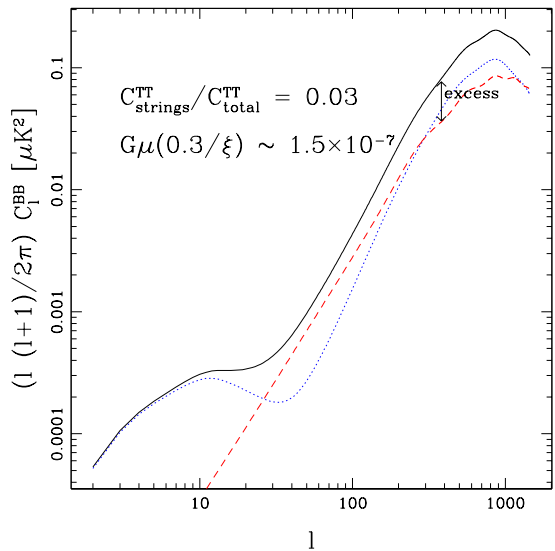} 
   \caption{A possible ``high"-$\ell$ detection scenario, in which a systematic excess of 
   B-mode power over what is expected from E to B lensing is observed. The black solid
   line is the sum of the string and lens-sourced B-mode power. The dotted blue line 
   is the string-sourced portion, while the red dashed line is the expected lensing spectrum.
   Here we have taken $G\mu (0.3/\xi) \simeq 1.5 \times 10^{-7}$.}
   \label{detection}
\end{figure}

In the ``high'' range, lensing of E-mode to B-mode polarization is expected to be the 
dominant source of BB power. However, this is also the range where  the string-sourced B-mode
spectrum peaks. Hence, we were able to do a calculation in this range similar to the one
described above to discover how much at what $G\mu$ strings would source the same
amount of B-mode power as lensing in this ``high'' range.
To calculate this, we summed the power in the $C_\ell^{BB}$s for $100 < \ell < 1000$. Our result is that the string sourced B-modes have the same power as lensed B-modes in the  ``high'' region
for $G\mu = 1.8 \times 10^{-7}$, assuming smooth strings. 
Again, this is a different kind of comparison than the one given above, in the ``low" region.
For ``low" $\ell$, it is quite possible that there will be no signal from gravity waves at all, and 
that a possible string signal will be the only source of power in this region. For ``high" $\ell$,
though, we will certainly find the lensed E to B-mode spectrum. 
So the proper approach in the ``high" range is to look for a systematic excess
in B-mode power over what is expected from lensing. This sort of excess is plotted in Fig. 
\ref{detection}, and was analyzed in some detail for the experiment C$\ell$over by \cite{Bevis:2007qz}.
 The fraction of TT power sourced by strings in this case would be only 3\%, yet
 B-mode signal in the ``high" region is doubled!  

For the experiments listed below, we designate each as {\sf L} if it covers the ``low'' range and {\sf H} 
if it covers the ``high'' range.

\begin{itemize}
\item {\sf QUaD} \cite{quad} ({\sf L/H}): This South Pole based bolometric array that has completed data collection. It explores the range $25 < \ell < 1000$, with sensitivity enough eventually to reach $r \geq 0.14 \rightarrow G\mu \geq 5.2 \times 10^{-7}$ in the ``low" range, and with
a good chance of seeing E to B lensing (and with it, a string-sourced excess) in the ``high" range. Results from its first season of operation were reported in \cite{Ade:2007ty}, while the full results
should be available soon \cite{carlstrom}.
\item {\sf BICEP} \cite{bicep} ({\sf L}): This experiment is presently taking its first season of data. Similar to QUaD, it is located at the South Pole and explores between  $20 < \ell < 100, \, 120$, with enough sensitivity to reach $r \geq 0.1 \rightarrow G\mu \geq 4.4 \times 10^{-7} $ by its third year (2009).
\item {\sf EBEX} \cite{ebex} ({\sf L/H}): A balloon-based experiment that is expected to begin taking data in 2008. It will observe in the range $20 < \ell < 1000$, with greatest sensitivity at high-$\ell$. Though its proposal is not explicit on this point \cite{ebex}, they appear to be able to observe $r \geq 0.06 \rightarrow G\mu \geq 3.4 \times 10^{-7}$ at low-$\ell$, and have enough sensitivity at higher $\ell$ that they should be able to detect E to B lensing, and hence observe a string-sourced excess
in the ``high" range. If $G\mu \gtrsim 3 \times 10^{-7}$, EBEX could definitively detect and confirm cosmic string-sourced B-mode polarization. 
\item C$\ell$over \cite{clover}({\sf L/H}): Located in the Atacama desert of Chile, C$\ell$over should begin taking data in 2009. It will explore between $20 < \ell < 1000$, and has enough sensitivity to see $r \geq 0.02 \rightarrow G\mu \geq 2 \times 10^{-7}$ in the low-$\ell$ band. Like EBEX, C$\ell$over should be able to detect E to B lensing, and so could detect and confirm a cosmic string signal down to $G\mu \geq 10^{-7}$. 
\item {\sf QUIET} \cite{quiet} ({\sf L/H}): Also to be located in the Atacama, QUIET operates in two $\ell$ ranges, and 
expects to operate through two operational phases. In phase one, 
it will observe at (1) $\ell \sim 100$, reaching at least $r = 0.18 \rightarrow G\mu = 6\times 10^{-7}$ and (2) between $500 < \ell < 1000$ with enough sensitivity to measure E to B lensing. In phase two, they hope to give best-of-breed measurements between $50 < \ell < 250$ and $450 < \ell < 1700$, though it is unclear what $r$ they hope to reach. Construction for this instrument is underway. 
\item{PolarBearR} \cite{polarbear} ({\sf H}): An experiment to be placed on the VIPER South Pole telescope, concentrating on the range $100 < \ell < 2000$. PolarBeaR-I, the first phase, should be able to detect $r=0.1$ in 
the low-$\ell$ range and give a very well characterized measurement of the E to B lensing signal up to $\ell \sim 900$,
with at least 9 $\ell$ bins in the {\sf H} range. This level of resolution in $\ell$ space could allow a cosmic
string signal to be differentiated from a lensing signal for $G\mu \gtrsim 10^{-7}$. A possible later upgrade, 
PolarBeaR-II, would give a very fine-grained ($\sim 20$ $\ell$-bins) measurement of the {\sf H} region.
\end{itemize}

\section{Conclusion}
\vspace{-0.1in}

Cosmic string networks can produce  B-mode polarization in the CMB at observable levels if the strings source
between 1 and 10\% of the TT-power in the CMB. This prediction holds across a wide variety of string network parameter values, as we have discussed above. 
Smooth ($\alpha \sim 1$) and/or slow-moving ($v_{rms} \sim 0.1$) strings produce the strongest B-modes, 
because they generate the largest ratio of anisotropic to isotropic stress sourced by cosmic strings.

The first B-mode polarization experiments will bin many $\ell$ multipoles together.
We calculate that a cosmic string-sourced B-mode signal
could mimic the B-mode power expected from inflationary gravity waves in the range between $2 < \ell < 100$.  
Since low-$\ell$ power from strings versus inflationary gravity waves scale approximately 
as $G\mu  \leftrightarrow 1.4 \times 10^{-6} \sqrt{r} \, \alpha \, ( 0.65 /v_{rms} ) \sqrt{\xi/0.3}$,
expectations that future experiments could probe $r \sim 10^{-3}$ \cite{Amarie:2005in}
 imply that cosmic strings with tensions as low as 
$5 \times 10^{-8}$ may eventually be detectable via their low-$\ell$ B-mode polarization.
In the meantime, without fine resolution in $\ell$, the only way to discriminate between these
two sources would be to look at high-$\ell$ ($100 < \ell < 1500$), where the string sourced B-mode spectrum
reaches its peak. In this range, gravitational lensing of E-mode polarization into B-mode polarization is expected to dominate, so a string contribution would appear as a systematic excess of power
over what is expected (see Fig. \ref{detection}).

Disentangling the string-sourced and primordial gravity wave signals will require either good spectral
information, via narrow $\ell$ bins, or accurate and reliable calculations of the
spectrum and amplitude of E to B lensing signal. 
Some existing methods for calibrating lensing, however, rely on the
assumption that all of the B-mode polarization present for $\ell > 150$ is sourced by lensing \cite{Seljak:2003pn}.
In our model, strings generate the same B-mode power as is expected from the E to B lensing signal for $G\mu \simeq 1.8 \times 10^{-7}$ (for standard network parameters and smooth strings); thus, strings with a tension in this range would lead to a doubling of the B-mode power in the ``high" region.

If low-$\ell$ B-modes or an excess of power in the high-$ell$ region are seen, we
should look for cosmic strings with tensions $G\mu\gtrsim 10^{-8}$ in another observational probe,
such as gravitational lensing \cite{Mack:2007ae, Chernoff:2007pd} or small-angle anisotropies in the CMB temperature \cite{Fraisse:2007nu, Amsel:2007ki}.
Such an observation would also motivate a reworking of lensing reconstruction algorithms to account for the cosmic string contribution.
Even in the absence of such analysis, 
detailed spectral information should potentially allow one to distinguish between
the inflationary and stringy reionization peaks at low-$\ell$.  In practice,
we expect that this will give only ambiguous results in the near term.
Accurately calibrating the E to B lensing spectrum and amplitude
is more promising, 
since it allows direct access to the peak in string-generated
B-mode power, which -- as noted before -- will be present as an apparent excess
of power in this high multipole range.

The discovery of a cosmic string signal in the B-mode polarization of the CMB would teach us about the string network that generated it since the shape of the string-sourced spectrum is 
sensitive to network parameters.
It would also help us to determine the energy scale at which inflation occurred.
As we have demonstrated, the expected string-sourced B-mode spectrum is robust
to theoretical uncertainty in the string network model. Thus, as these experiments turn on and collect
polarization data, a cosmic string network will either be seen through its polarization spectrum
or will be restricted to a considerably lower tension in the case of non-observation, with
polarization-based limits on $G\mu$ being comparable to those expected from the
analysis of the data from PLANCK \cite{Battye:2007si}.

\vspace*{-0.35in}
\acknowledgments
We acknowledge discussions with participants of the ``Cosmology and Strings'' workshop at the Abdus Salam International Center for Theoretical Physics, especially Mark Hindmarsh and Renata Kallosh, that helped to stimulate this work. We also thank Amber Miller and John Carlstrom for helpful discussions.
The work of L.~P. is supported by the National Science and Engineering Research Council of Canada (NSERC). The work of M.~W. at the Perimeter Institute is supported by the Government
of Canada through Industry Canada and by the Province of Ontario through
the Ministry of Research \& Innovation.

\end{document}